\documentclass[oldversion]{aa}  
\usepackage{graphicx}
\usepackage{txfonts}
\begin{document}

   \title{The eccentricity-mass distribution of exoplanets: signatures of
different formation mechanisms?}
   \titlerunning{The eccentricity-mass distribution of exoplanets}

   \author{Ignasi Ribas\inst{1}
          \and
          Jordi Miralda-Escud\'e\inst{1,2}
          }

   \offprints{I. Ribas}

   \institute{Institut de Ci\`encies de l'Espai (CSIC-IEEC), Campus UAB,
              Facultat de Ci\`encies, Torre C5 - parell - 2a planta, 
              08193 Bellaterra, Spain\\
              \email{iribas@ieec.uab.es, miralda@ieec.uab.es}
         \and
              Instituci\'o Catalana de Recerca i Estudis Avan\c cats,
              Barcelona, Spain\\
             }

   \date{}

\abstract{
We examine the distributions of eccentricity and host star metallicity of 
exoplanets as a function of their mass. Planets with $M \sin i \gtrsim 4$ 
M$_{\rm J}$ have an eccentricity distribution consistent with that of 
binary stars, while planets with $M \sin i \lesssim 4$ M$_{\rm J}$ are 
less eccentric than binary stars and more massive planets. In addition, 
host star metallicities decrease with planet mass. The statistical 
significance of both of these trends is only marginal with the present 
sample of exoplanets. To account for these trends, we hypothesize that 
there are two populations of gaseous planets: the low-mass population 
forms by gas accretion onto a rock-ice core in a circumstellar disk and is 
more abundant at high metallicities, and the high-mass population forms 
directly by fragmentation of a pre-stellar cloud. Planets of the first 
population form in initially circular orbits and grow their eccentricities 
later, and may have a mass upper limit from the total mass of the disk 
that can be accreted by the core. The second population may have a mass 
lower limit resulting from opacity-limited fragmentation. This would 
roughly divide the two populations in mass, although they would likely 
overlap over some mass range. If most objects in the second population 
form before the pre-stellar cloud becomes highly opaque, they would have 
to be initially located in orbits larger than $\sim 30$ AU, and would need 
to migrate to the much smaller orbits in which they are observed. The 
higher mean orbital eccentricity of the second population might be caused 
by the larger required intervals of radial migration, and the brown dwarf 
desert might be due to the inability of high-mass brown dwarfs to migrate 
inwards sufficiently in radius. 
}


   \maketitle

\section{Introduction}

The discovery of the first exoplanets around main-sequence stars a decade 
ago (Mayor \& Queloz 1995; Butler \& Marcy 1996; Marcy \& Butler 1996; 
Butler et al. 1997) has provided us with a first glimpse at the 
fascinating diversity of planetary systems in the universe. At present, 
with about 200 exoplanets known (most of them having been found by the 
method of radial velocities around stars similar to the Sun, and a few by 
the methods of transits and microlensing events), we have learned that: 
{\em (a)} Jovian planets are found over a wide range of orbital radii $a$, 
from the smallest orbits where they can survive evaporation ($a \sim 0.03$
AU) to the largest ones at which they can be detected ($a\sim 5$ AU); 
{\em (b)} Jovian planets that have not been tidally circularized show a 
wide range of eccentricities, with a median value of $\sim 0.3$; {\em (c)} 
orbital resonances are commonly found in systems when more than one planet 
is detected; {\em (d)} there is a power-law distribution of planet masses 
that is roughly uniform in $\log M$ from $\sim 0.3$ to $\sim10$ Jupiter 
masses (e.g., Tabachnik \& Tremaine 2002), although with a strong deficit 
of objects with mass greater than $\sim0.015$ M$_\odot$ (designated as 
brown dwarfs), which is known as the ``brown dwarf desert'', and probably 
an increase in the planet abundance at lower masses, as suggested by the 
recent detections of Neptune-like planets with the method of microlensing 
(Beaulieu et al. 2006; Gould et al. 2006). All of these findings have come 
as surprises, and none of them was predicted or expected from theories of 
planetary formation.

With the increasing sample of exoplanets available for statistical 
studies, the fundamental question of the planet formation mechanism and 
their subsequent orbital evolution that results in the observed 
distribution may start to be addressed. The only information we have so 
far that relates to the formation process of planets is the distribution 
of their orbital periods, eccentricities, and planet masses, which can 
also be Compared to the same distributions for binary stars and brown 
dwarfs. In addition, properties of the host stars such as their 
metallicity can be included. Jovian planets are thought to form in 
circumstellar disks from the coalescence of planetesimals and 
gravitational accretion of gaseous material (Pollack et al.\ 1996). This 
process should lead to orbits that are initially circular, but they could 
subsequently be perturbed by several mechanisms causing an increase of the 
eccentricity, such as disk-planet interactions (Kley \& Dirksen 2006), the 
Kozai mechanism (Kozai 1962; Holman et al. 1997; Takeda \& Rasio 2005), 
and close encounters or resonant interactions between planets (Chiang et 
al. 2002; Ford et al. 2005; see also the review on eccentricity growth 
mechanisms by Tremaine \& Zakamska 2004). These mechanisms need to be 
effective for a large majority of planets formed in disks since most of 
the detected exoplanets are found to possess eccentricities much larger 
than solar system planets (and hence, the solar system must be an oddity 
among planetary systems).

It has been noted before that the distribution of eccentricities of the 
exoplanets seems to be remarkably similar to that of binary systems 
(Heacox 1999; Stepinski \& Black 2000, 2001), with a slight tendency for 
the eccentricities to increase with planet mass (see Tremaine \& Zakamska 
2004; Marcy et al. 2005; Papaloizou \& Terquem 2006). This result is 
surprising because it is not clear how a dynamical process can result in 
higher eccentricities acquired by more massive planets. Although 
interactions with a gaseous disk can generate eccentricities more easily 
for the most massive planets, it is doubtful that this process alone can 
excite the eccentricities up to the observed distribution. While the Kozai 
mechanism predicts an eccentricity distribution independent of planetary 
mass, perturbations among planets would likely tend to leave lower mass 
planets with higher eccentricities, basically from energy equipartition 
arguments; it is possible, however, that if one observes the most massive 
planet that has survived in every planetary system, the systems with 
greater total mass compared to the host star have been more strongly 
perturbed.

Contrary to planetary objects, the most favored hypothesis to explain the 
formation of binary stars and brown dwarf companions is by fragmentation 
of the parent molecular cloud during the gravitational collapse process, 
or as a result of gravitational instability or fission of a rapidly 
rotating pre-stellar cloud (Tohline 2002). If the fragmentation process 
resulted in an orbit randomly selected from phase space for a fixed 
orbital energy, the distribution of eccentricities should be uniform in 
$e^2$, with a median eccentricity of $0.7$ in a binary star sample. 
However, this is not observed (e.g., Abt 2005), and even after removing 
systems that may have been affected by tidal circularization the typical 
eccentricities are substantially lower. It is particularly striking that 
the distribution of eccentricities of exoplanets and binary stars are 
remarkably similar, since the two types of systems are thought to form by 
different mechanisms.
  
In this paper, we consider the possibility that some planetary mass 
objects have also formed by the process of direct cloud fragmentation. The 
known exoplanets would therefore be part of two populations (as proposed 
earlier by, e.g., Black 1997; Mayor et al. 1998; Papaloizou \& Terquem 
2001; Udry et al. 2002), stemming from two different mechanisms of mass 
growth: either gas accretion from a disk onto a seed planetesimal, or 
fragmentation during the collapse of the gas cloud. Even though exoplanets 
and brown dwarfs are usually considered as separate classes of objects 
(with the separation chosen at a mass of $0.013$ M$_\odot$, the minimum 
mass required for deuterium burning), there is no fundamental reason why 
these objects should belong to distinct classes from the point of view of 
their formation. In general, the different formation processes mentioned 
above (as well as disk fragmentation by gravitational instability) may be 
relevant for objects over wide ranges of mass that may overlap, and should 
be unrelated to the minimum mass for deuterium burning or for hydrogen 
burning. We shall search for possible signatures of the presence of two 
populations of objects among the known exoplanets and brown dwarfs with 
the improved statistics available today, and also examine the correlation 
with the metallicity of the host star.

\section{Eccentricity distribution of exoplanets and binary stars}
\label{sec_ecdis}

We start by examining the distribution of eccentricity and $M \sin i$ in 
the sample of known exoplanets and brown dwarf companions within 5 AU 
of their star. We use the sample of all known exoplanets around normal 
stars with measured orbital elements from radial velocities (see, e.g., 
the Extrasolar Planets Encyclopaedia; Butler et al. 2006). We add to these 
a set of 17 objects with $0.013$ M$_\odot$ $< M \sin i <$ 0.08 
M$_\odot$, which are within 5 AU of a star of F, G, or K spectral type 
(note that more brown dwarfs are known at this distance from M dwarfs, 
where they seem to be more common; see Close 
et al. 2003). This forms our sample of ``substellar objects'', a total of 
204. Of course, many of the 17 objects with $M \sin i > 0.013$ M$_\odot$ 
may actually be stars. Because of the presence of the brown dwarf desert 
(a dearth of brown dwarfs orbiting within 5 AU of a solar-type star 
compared to either stellar or planetary-mass companions), an object found 
with $M \sin i$ in the range corresponding to a brown dwarf mass is likely 
to be a star with a small orbital inclination. In fact, 6 of these 17 
objects (\object{HD 112758}, \object{HD 110833}, \object{HD 169822}, 
\object{HD 217580}, \object{HD 18445}, and \object{BD-04 782}) have 
measured astrometric orbits by Hipparcos that confidently place their 
masses in the stellar regime (as long as the error bars are correct; see 
Halbwachs et al. 2000; Vogt et al. 2002), another 3 are uncertain 
(\object{HD 283750}, \object{HD 114762}, and \object{HD 140913}; Halbwachs 
et al. 2000), and the other 8 are likely to be true brown dwarfs 
(\object{HD 180777}, \object{HD 89707}, \object{HD 137510}, \object{HD 
127506}, \object{HD 184860}, \object{HD 202206b}, \object{HD 168443c}, and 
\object{HD 29587}; see Halbwachs et al. 2000; Vogt et al. 2002; Endl et 
al. 2004; Galland et al. 2006). Although a brown dwarf desert is clearly 
present, it is not completely empty.

\begin{figure}
\centering
\includegraphics[width=8.8cm]{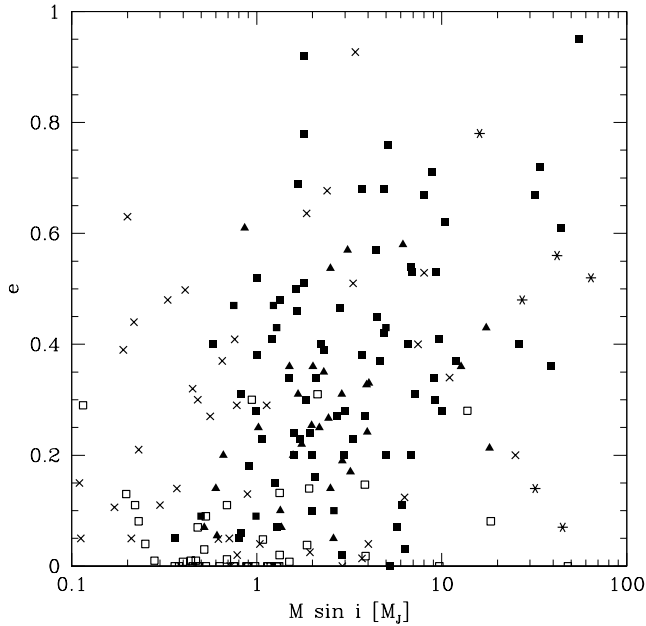}
   \caption{Eccentricity versus mass for the sample of substellar objects. 
{\it Open squares:} planets with $a < 0.1$ AU. {\it Crosses:} planets with 
$0.1$ AU $< a < 0.5$ AU. {\it Filled squares:} planets with $a > 0.5$ AU. 
{\it Triangles:} Planets with $a > 0.5$ AU in systems with more than one 
detected planet. {\it Stars:} Objects that have been found to have a 
stellar mass from Hipparcos astrometric observations.}
      \label{fig_ecdis}
\end{figure}

We show the distribution of eccentricity and $M \sin i$ for all the 
objects in our sample, 204 objects in total, in Fig. \ref{fig_ecdis} (10 
of the planets have $M \sin i < 0.1$ M$_{\rm J}$ and do not appear in the 
figure). We have plotted them with different symbols depending on their 
semimajor axis. Objects with $a < 0.1$ AU have likely been affected by 
tidal circularization, and clearly have small eccentricities. Some of the 
objects in the middle group could have been affected by tidal 
circularization, especially for $a< 0.2$ AU and very high eccentricities, 
and in cases where the planets remained larger than their present size for 
a substantial time in their youth. There is, however, no clear tendency 
for objects with $0.1$ AU $< a < 0.5$ AU to show smaller eccentricities 
than objects with $a > 0.5$ AU at similar values of $M \sin i$. Objects 
that are also at $a > 0.5$ AU and belong to systems with more than one 
planet detected do not have an obviously different distribution, 
suggesting that being part of a planetary system does not greatly affect 
the eccentricity. Finally, six objects with $M \sin i > 0.013$ M$_\odot$ 
are stars with low orbital inclinations according to Hipparcos 
measurements.

\begin{figure}
\centering
\includegraphics[width=8.8cm]{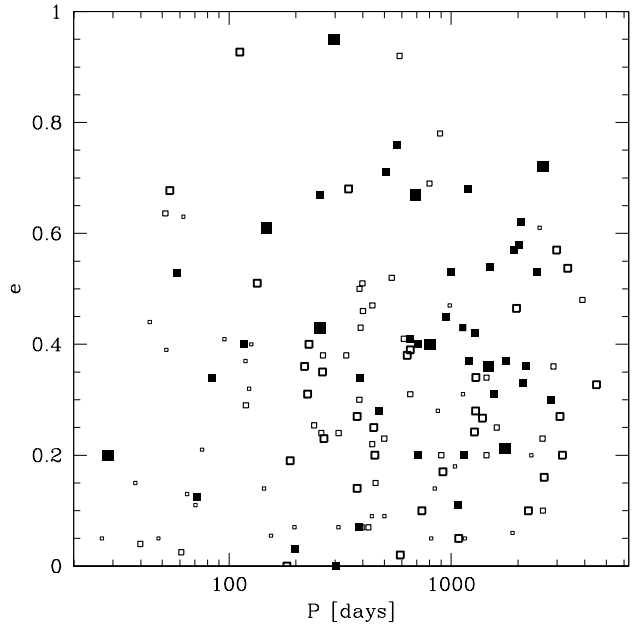}
   \caption{Eccentricity versus orbital period for all planets and brown 
dwarfs of different mass groups; objects with $a < 0.2$ AU have been 
removed from the sample. {\it Small open squares:} $M \sin i <$ M$_{\rm 
J}$; {\it medium open squares:} M$_{\rm J} < M \sin i < 2$ M$_{\rm J}$; 
{\it large open squares:} $2$ M$_{\rm J} < M \sin i < 4$ M$_{\rm J}$; {\it 
large filled squares:} $4$ M$_{\rm J} < M \sin i < 13$ M$_{\rm J}$; {\it 
extra-large filled squares:} $13$ M$_{\rm J} < M \sin i$. }
      \label{fig_ecp}
\end{figure}

One can also appreciate from Fig. \ref{fig_ecdis} that there is a tendency 
for the orbital eccentricity to increase with mass, as mentioned in the 
introduction. The effect is, however, a weak one (we have included the two 
recently discovered planets by Jones et al.\ 2006, which go against this 
mean tendency). There is a natural concern that this correlation might be 
induced by observational selection effects: according to Cumming (2004), 
high eccentricities tend to be more easily detectable for long period 
orbits, and low eccentricities are easier to detect for short periods. 
Because low-mass planets are harder to detect for long periods (because of 
the low velocity amplitude), a change in the eccentricity distribution 
with orbital period due to selection bias might induce the observed 
dependence of eccentricities with planet mass. To check for this 
possibility, in Fig.\ \ref{fig_ecp} we plot the eccentricity versus the 
period for the sample of all the exoplanets and brown dwarfs that have $a 
> 0.2$ AU (to remove the objects affected by tidal circularization), using 
different symbols for different ranges of $M \sin i$ (the largest points 
correspond to more massive planets). While one can also discern in Fig. 
\ref{fig_ecp} the tendency for eccentricity to increase with mass, there 
is no obvious variation of eccentricity with period for planets of fixed 
mass. We shall assume in this paper that the mass-eccentricity relation is 
not being severely affected by selection effects, although this will 
require more careful examination as the number of known exoplanets 
increases.

We also use the sample of binary stars based on The 9$^{\rm th}$ Catalog 
of Spectroscopic Binary Orbits (Pourbaix et al. 2004). From this catalog 
we selected binary stars according to several criteria: {\em (a)} A
quality flag of 2 or greater to ensure reasonably firm 
orbital solutions and reliable eccentricities (tests using different 
selections in the quality flag indicate that this does not introduce any 
bias in the eccentricity distribution); {\em (b)} the orbital period is 
required to be larger than 30 days, large enough so that circularization 
processes have not played an important role; {\em (c)} only main-sequence 
or subgiant components are kept to minimize the range of stellar radii and 
permit the use of the orbital period as a measure of the significance of 
circularization processes. Similar criteria were recently used by Abt 
(2005) to study the eccentricity distribution of binary stars. The 
resulting sample is composed of 200 spectroscopic binaries meeting the 
restrictions described above. In addition, we consider the subsample of 
these 200 binaries having star components of FGKM spectral type, for 
similarity with exoplanet host stars. This subsample is composed of 130 
spectroscopic binaries.

\begin{figure}
\centering
\includegraphics[width=7.5cm]{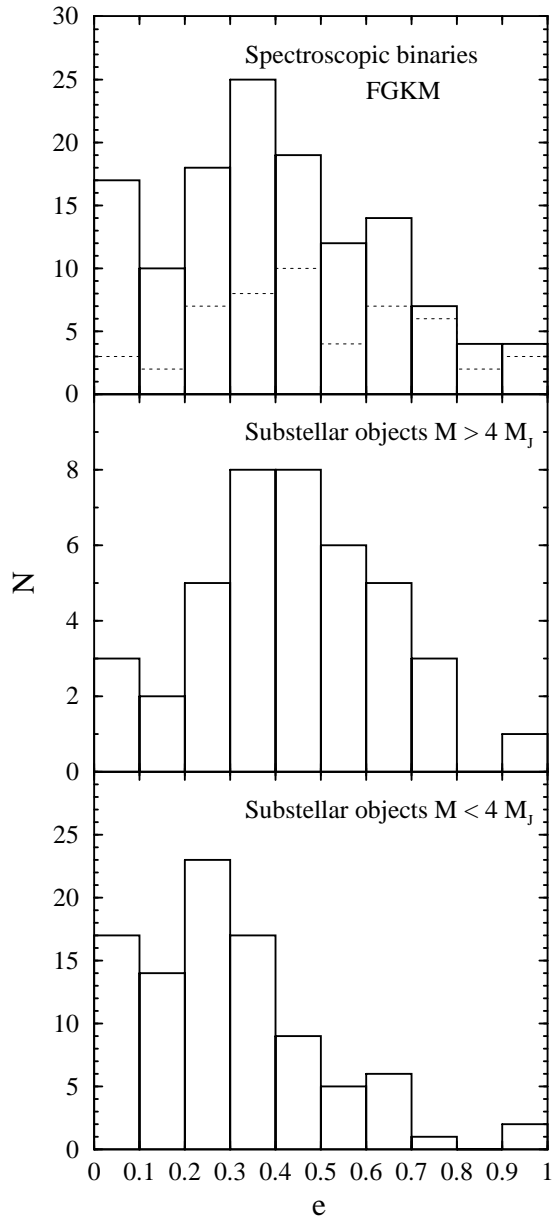}
   \caption{Eccentricity distributions for the three samples considered in 
this work: spectroscopic binaries (top), substellar objects with $M>4$ 
M$_{\rm J}$ (middle), and substellar objects with $M<4$ M$_{\rm J}$ 
(bottom). All objects with $a < 0.2$ AU have been removed from the sample 
to avoid the effects of tidal circularization. The histogram represented 
with a dotted line in the top panel corresponds to double-lined 
spectroscopic binaries only (see text for discussion).}
      \label{fig_hist}
\end{figure}

The eccentricity distributions of the samples considered in this work are 
shown as histograms in Fig. \ref{fig_hist}. The top panel is for the 
spectroscopic binaries with components of FGKM spectral types. The middle 
and bottom panels show the eccentricity distributions of substellar 
objects with minimum masses above and below 4 M$_{\rm J}$, respectively 
(the middle panel includes the 8 objects with $13$ M$_{\rm J} < M \sin i < 
80$ M$_{\rm J}$ that we consider to be likely brown dwarf companions, as 
discussed above). The placing of the mass division at 4 M$_{\rm J}$ is 
somewhat arbitrary at this point, although there are observational 
indications of a change in planet properties around this value (Udry et 
al. 2002). Further discussion on the mass limit is provided in Sect. 
\ref{sec_disc}. All substellar objects with $a < 0.2$ AU have been 
eliminated from the sample to remove any orbits that have been affected by 
tidal circularization.

To compare the distributions we made use of the Kolmogorov-Smirnov (K-S) 
two-sample test following the implementation of Press et al. (1992). The 
tests were carried out with the three samples described above. Given two 
cumulative distribution functions $S_1(x)$ and $S_2(x)$ with $n_1$ and 
$n_2$ data points, respectively, the K-S statistic is defined as 
$D=\mbox{max~}|S_1(x)-S_2(x)|$ and the effective number of data points is 
$n_{\rm eff}=(n_1 n_2)/(n_1+n_2)$. Then the probability that the real 
value of $D$ is greater than its observed value $D_{\rm obs}$, $p(D>D_{\rm 
obs})$, yields an estimate of the likelihood of the null hypothesis that 
the two distributions have been drawn from the same population.

The results of the K-S tests are presented in Table \ref{tab_kstest}. The 
sample of spectroscopic binaries and high-mass planets clearly have 
eccentricity distributions consistent with being identical. However, the 
eccentricity distribution of the low-mass planets is found to be 
different. The probability for the distributions to be the same is around 
0.1\% to 0.2\%. The main difference in the distributions is that low-mass 
planets tend to have lower eccentricities, as shown in Fig. 
\ref{fig_hist}.

\begin{table}
\caption{Results of the Kolmogorov-Smirnov tests}
\label{tab_kstest}
\centering
\begin{tabular}{c c c c c}
\hline\hline
Distrib \#1 &
Distrib \#2 &
$D_{\rm obs}$&
$p(D>D_{\rm obs})$&
$n_{\rm eff}$ \\
\hline
$M \sin i<4$ M$_{\rm J}$ & $M \sin i>4$ M$_{\rm J}$ & 0.359 & 0.001 & 28 \\
SB FGKM                  & $M \sin i<4$ M$_{\rm J}$ & 0.251 & 0.002 & 54 \\
SB FGKM                  & $M \sin i>4$ M$_{\rm J}$ & 0.125 & 0.682 & 31 \\
SB all                   & $M \sin i<4$ M$_{\rm J}$ & 0.238 & 0.001 & 63 \\
SB all                   & $M \sin i>4$ M$_{\rm J}$ & 0.150 & 0.400 & 34 \\
\hline
\end{tabular}
\end{table}

\subsection{Binary stars on nearly circular orbits}

There is an additional surprising feature in the eccentricity distribution 
of the spectroscopic binaries: the excess of binaries with very low 
eccentricities, $e < 0.1$, compared to a distribution that grows linearly 
with $e$ for small $e$, as expected if all orbits have been subject to a 
similar dynamical evolution inducing random variations in their 
eccentricity, which should uniformly fill the available phase space at 
small $e$. We have checked that this excess is not the result of systems 
with unreliable observations assumed to have a circular orbit, and it does 
not appear to be due to any other observational bias.
 
To further investigate this excess of low-eccentricity orbits, the dotted 
histogram in the top panel of Fig. \ref{fig_hist} shows the distribution 
of double-lined spectroscopic binaries only. Double-lined binaries 
generally have components of similar mass. Very roughly, a system in which 
the ratio of luminosity of the two components is above 0.1--0.05 will 
permit the identification of both lines in the spectra, and this ratio of 
luminosities corresponds to a mass ratio $\sim 0.5$ for main-sequence 
components. We can see that the excess of low-eccentricity orbits is more 
pronounced in systems with unequal components.

This feature in the eccentricity distribution suggests that a small 
fraction ($\sim$ 10\%) of the binaries are formed on nearly circular 
orbits and experience very little orbital evolution afterwards. High-mass 
planets might also have this excess of orbits at small eccentricities, 
although their numbers are at present too small to make this statistically 
significant. Low-mass planets do not seem to show any excess of 
low-eccentricity orbits.

\section{Host star metallicity distribution of exoplanets}

Another property of exoplanets that can inform us about their formation 
mechanism is the metallicity of the host star. The probability that a star 
hosts planets of the type detected in radial velocity surveys increases 
rapidly with stellar metallicity (e.g., Fischer \& Valenti 2005). The 
implication is that the formation of Jovian planets is made more likely 
when heavy elements have a high abundance (the possibility that stellar 
metallicities at the photosphere are increased by accreted planets was 
discarded by examining the dependence of the metallicity with the spectral 
type of the star; see Pinsonneault et al. 2001). Our examination of the 
eccentricity--planet mass correlation in Sect. \ref{sec_ecdis} suggested 
that planets may be divided into two populations, roughly those more and 
less massive than $\sim4$ M$_{\rm J}$. It is therefore worthy to see if 
the metallicity of the host star shows any similar change around a 
comparable mass.

\begin{figure}
\centering
\includegraphics[width=8.8cm]{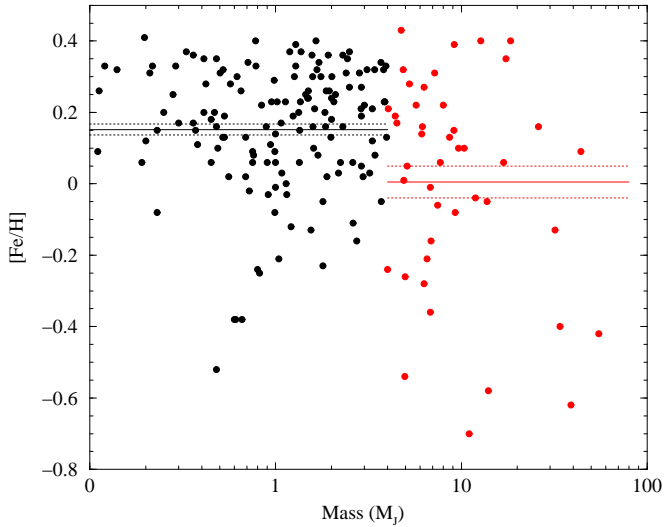}
   \caption{Mass versus metallicity for the substellar objects in the 
sample. Mean metallicity values for the two proposed populations (minimum 
mass above and below 4 M$_{\rm J}$) are represented by the horizontal 
lines (with the corresponding error bars shown as dotted lines).}
      \label{fig_met}
\end{figure}

Figure \ref{fig_met} shows the metallicities of all stars hosting planets, 
collected from the literature, as a function of the planet mass. The 
average host star metallicity of planets with $M\sin i < 4$ M$_{\rm J}$ 
and $M \sin i > 4$ M$_{\rm J}$ is indicated as is the error of the average 
metallicity. The average metallicity of stars with planets of mass below 4 
M$_{\rm J}$ is $[Fe/H]=0.152\pm0.015$, while stars with substellar objects 
above that mass have a mean metallicity of $[Fe/H]=0.005\pm0.045$. The 
difference is therefore at the 3-$\sigma$ level. For comparison, the 
average metallicity in the solar neighborhood is $[Fe/H]\simeq -0.15$ with 
a scatter of $\sim 0.2$ dex (Nordstr\"om et al.\ 2004; Fischer \& Valenti 
2005; Valenti \& Fischer 2005). The population of stars with massive 
substellar objects probably has a metallicity value consistent with that 
of the field, especially when considering that the sample of stars for 
which radial velocity searches have been made is biased to high 
metallicity. Visual examination of the dots suggests that the tendency of 
the host star metallicity to decrease with planet mass is a gradual one, 
over the range from $\sim 1$ to $50$ M$_{\rm J}$.

\section{Discussion} \label{sec_disc}

\subsection{A hypothesis for the physical origin of two planet
populations}

The statistical evidence for the two trends we have examined, of 
increasing orbital eccentricity and decreasing host star metallicity with 
planet mass, should be considered as marginal at this point. However, the 
presence of the two independent trends reinforce each other in suggesting 
that there may be different planetary formation mechanisms that give rise 
to planets of different masses. In this section, we discuss a possible 
physical explanation for these trends, assuming that they are correct and 
bearing in mind that they will need to be confirmed (or refuted) by 
improved statistical evidence as the number of known exoplanets increases.

The metallicity of the host star should not be affected by the fact that a 
planet has formed and has stayed in orbit around the star. Therefore, it 
seems reasonable to conclude that the majority of the high-mass planets 
form in a way that is less influenced by the metallicity than low-mass 
planets. The most natural interpretation is that many of the high-mass 
planets form through the same mechanism postulated for brown dwarfs, by 
direct fragmentation of the pre-stellar cloud (possibly combined with the 
interruption of gas accretion by radiation feedback from nearby stars) or 
of a circumstellar gaseous disk that is gravitationally unstable (e.g., 
Whitworth et al. 2006), while the low-mass planets form by the initial 
coagulation of a core of rock and ice and subsequent gas accretion onto 
the core (Pollack et al. 1996). There is indeed no reason to suppose that 
substellar objects formed by these two mechanisms would not overlap in 
mass. The formation of a planet from a core of rock and ice has a natural 
mass upper limit determined by the total mass that can be accreted from a 
circumstellar disk, at the time and the radius where the planetary core 
reaches a critical mass allowing gas accretion to start. At the same time, 
a natural lower limit to the mass of objects formed by fragmentation and 
direct gravitational collapse of gas is the opacity limit, which assumes 
that fragmentation will not be efficient once a pre-stellar cloud becomes 
opaque to its own cooling radiation, and is given by (Rees 1976): 
\begin{equation} M_{\rm op}= M_{\rm Ch}\, \mu^{-9/4}\, f^{-1/2}\, \left( { 
kT \over m_p c^2 } \right)^{1/4} ~, \label{opam} \end{equation} where 
$M_{\rm Ch}$ is the Chandrasekhar mass, $\mu$ is the mean molecular weight 
in units of the proton mass, $f$ is the emissivity of the cloud at the 
moment it becomes opaque (expressed as the fraction of the blackbody 
radiation that is emitted), and $T\sim 10$ K is the cloud temperature. 
This opacity limit mass is likely to represent the smallest possible mass 
of an object formed by direct gravitational collapse of gas: even if 
fragmentation is still possible by gravitational instability in a rapidly 
rotating, highly opaque disk, it is unlikely that the gas temperature in a 
circumstellar disk around a young star would be low enough to bring the 
Jeans mass below the opacity limit value in the conditions of the 
protostellar nebula. The opacity limit mass is close to a Jupiter mass, 
suggesting that objects formed by gas fragmentation may extend into the 
high-mass planet regime and be found orbiting other stars with an 
abundance that is independent of the host star metallicity. On the other 
hand, gas planets formed by core accretion may not usually grow to more 
than a few Jupiter masses, form preferentially around metal-rich host 
stars starting from nearly circular orbits, and grow their orbital 
eccentricities to an average value lower than the more massive objects 
formed by fragmentation.
  
\subsection{Orbital migration of the high-mass population}

However, this simple idea cannot account for the observations without 
including an additional ingredient. If the lower limit to the mass of 
substellar objects formed by fragmentation of gas clouds is to have 
anything to do with the opacity limit, then these objects should form at 
very large distances from their stars because the density at which opacity 
sets in is $n_H \sim 10^{10}$ cm$^{-3}$ (e.g., Whitworth et al. 2006), and 
the size of a region containing the mass in Eq. (\ref{opam}) at this 
density is $\sim 30$ AU. Therefore, these objects would have to be formed 
at large distances and then migrate to the much smaller orbits at which 
the known exoplanets have been detected.

It may not be unreasonable for the population of high-mass planets to have 
undergone a large radial migration. After all, we know that it is 
necessary for most of the planets to have experienced radial migration 
because it is believed that they can form only at radii large enough to 
allow for the presence of ice particles, and many planets are found within 
the ice-line. If planets always start on circular orbits when forming from 
a rock-ice core, their eccentricities would have to increase as they 
experience radial migration to account for the observed high 
eccentricities. The need for a large interval of radial migration for 
objects formed by fragmentation might then be the reason behind the 
increasing eccentricity with mass: while most of the high-mass planets 
would have migrated over large intervals, some of the low-mass planets, 
which could form from solid cores in the protoplanetary disk at distances 
much closer than the high-mass population, may have migrated over radial 
intervals that are too small to grow the eccentricity by the same amount.

Possible mechanisms for the migration of planets formed at large distances 
include the dynamical interaction with low-mass planets formed in the 
circumstellar disk, a hydrodynamic and/or gravitational interaction with 
the gaseous circumstellar disk, or interaction with a disk of 
planetesimals. The orbit of a distant, high-mass planet could be perturbed 
to a small pericenter by external torques, interactions with other distant 
planets, or a Kozai mechanism, and then its semimajor axis could be 
greatly reduced by exchanging energy with planets formed at short 
distances, or by dissipative hydrodynamic or gravitational interactions 
with the disk.

If brown dwarfs and high-mass planets in orbit around solar-type stars are 
indeed formed only at large distances, this might account also for the 
presence of the brown dwarf desert (note that brown dwarfs are abundant 
around M-type stars at typical distances of $\sim$ 1 AU, where they have 
probably formed by a mechanism similar to binary stars; Close et al. 2003). 
As the brown dwarf mass increases, a more massive disk needs to be present 
around the star in order that a sufficient amount of angular momentum can 
be absorbed by the disk to allow for the migration of the brown dwarf. It 
is possible that planets or brown dwarfs above some critical mass are not 
typically able to migrate to within a few AU of their host star, because 
their angular momentum is too large to be exchanged with the circumstellar 
disk.

Clearly, a model where high-mass planets and brown dwarfs form only at 
large distances predicts that many of these objects should still be found 
at late times in orbits of large radius around stars. If $\sim$ 1\% of 
solar-type stars contain high-mass planets within 5 AU, then a similar 
number of stars should contain the same planets and brown dwarfs at 
distances $>$ 30 AU, even if the migration process of these distant 
planets can be highly efficient (that is, a high fraction of the distant 
planets end up in orbits within 5 AU). In general, brown dwarfs and 
planets with $M> 5 {\rm M}_{\rm J}$ are abundant in young clusters, 
accounting for $\sim$ 10\% of all the objects (see Gonz\'alez-Garc\'\i a 
et al. 2006, although in some cases the abundance may be lower, see Lyo 
et al. 2006). These same objects in young clusters are found only rarely as 
companions to solar-type stars on distant orbits (in $\sim$ 1\% of the 
stars; R. Rebolo, priv. communication), although the abundance of these 
companions could be lower in clusters containing many stars compared to 
the field, because of tidal destruction of the binary systems. The case of 
the star $\epsilon$ Indi is noteworthy: at a distance of 3.6 parsecs, it 
is the 20th nearest star to the Sun and it has a pair of brown dwarfs at a 
separation of 1500 AU that was discovered only very recently (McCaughrean 
et al. 2004). In summary, it seems plausible that there is a large enough 
reservoir of high-mass planets and brown dwarfs formed at large distances, 
some of which may undergo radial migration and account for the high-mass 
planets found within 5 AU of their stars. The alternative possibility to 
the migration scenario would be that the high-mass planets can form at a 
radius near their final orbit at which they are observed, by direct gas 
collapse through gravitational instability of a rotating, opaque disk. 
However, if this mode of formation is to account for both binary stars and 
high-mass planets, the presence of the brown dwarf desert has no clear 
explanation.
  
\subsection{The eccentricity distribution of planets and binary stars}

A possible test for the idea that the large eccentricities of the 
high-mass planet population originate in a large interval of radial 
migration is related to the fact that all planets on small orbits should 
have migrated by large intervals, because planets formed by the core 
accretion process are thought to have formed in orbits outside the 
ice-line, so they can reach small orbits only by migrating. Therefore, at 
a small semimajor axis the eccentricity distribution should become more 
alike for low- and high-mass planets. However, it is hard to tell from 
Fig. 2 if there is any tendency of increasing eccentricity with decreasing 
period, and as mentioned in Sect. \ref{sec_ecdis} any such dependence 
might be influenced by selection effects arising from the methods by which 
planets are found in Doppler surveys.
  
The hypothesis we have examined for the origin of the population of 
high-mass planets provides no satisfying explanation for why the 
eccentricity distributions of spectroscopic binaries and high-mass planets 
are so similar. Clearly, binary stars would have even greater difficulty 
for migrating inwards from a large orbit than brown dwarf companions. 
Binary stars are likely to form on an orbit of similar size to their final 
one, perhaps by gravitational instabilities in massive circumstellar disks 
or fission of a rapidly rotating pre-stellar cloud (see Tohline 2002 for a 
review of binary formation theories). In any case, it is clear that binary 
stars must form by very different mechanisms than any high-mass planets 
collapsing directly out of gas. Even if the high-mass planets formed on 
orbits of similar size as the binary stars, the need to account for the 
brown dwarf desert strongly suggests totally different formation 
mechanisms. Binary stars are likely to form in different ways in any case, 
in view of the wide range of orbital sizes, the known excess of twin 
binaries (with very similar component masses; Pinsonneault \& Stanek 
2006), and the excess of small eccentricity orbits we mentioned at the end 
of Sect. \ref{sec_ecdis}. The similar eccentricity distributions of binary 
stars and high-mass planets may more likely be related to processes that 
occur after their formation; for example, migration of a high-mass planet 
through a disk and the formation of a star from gravitational instability 
in the disk might give rise to similar eccentricity distributions at the 
end of the process. In addition, a series of perturbations that change 
orbital eccentricities in a random way may typically produce a 
characteristic eccentricity distribution (see Juric \& Tremaine 2006).

\section{Summary}

We have found that the known exoplanets exhibit two trends in their 
properties that, although statistically marginal at this point, may be 
indicative of the presence of more than one population formed by different 
mechanisms: {\em (a)} The mean orbital eccentricity tends to increase with 
planet mass, and {\em (b)} the metallicity of the host star tends to 
decrease with planet mass. We also find that {\em (c)} the eccentricity 
distributions of planets more massive than a few Jupiter masses and of 
spectroscopic binaries are remarkably similar. Other known relevant facts 
for understanding the origin of exoplanets include the following: {\em 
(d)} many exoplanets must have migrated over large radial intervals to 
reach their present orbits from the location where they could form, {\em 
(e)} their final orbits must have been left with the observed average 
eccentricity $\bar e \sim 0.3$ (for planets that have not undergone tidal 
circularization) at the end of this migration process, {\em (f)} there is 
a brown dwarf desert in the mass distribution of orbiting objects that 
includes stellar and substellar companions.

We favor the hypothesis whereby the exoplanets that are being discovered 
at present, mostly in radial velocity surveys, actually constitute two 
different populations that overlap in mass. The first population forms by 
the initial assembly of a core of rock and ice, with subsequent gas 
accretion from a disk, and is found more frequently in high metallicity 
stars. The second population forms by direct collapse of gas, and its 
abundance is independent of stellar metallicity. Under the additional 
hypothesis that this second population forms by fragmentation before the 
gas cloud becomes opaque (rather than from gravitational instability in a 
rapidly rotating, highly opaque pre-stellar cloud), we suggest that the 
opacity limit is a natural lower limit to the mass of this class of 
objects, and that all the observed high-mass planets within 5 AU of 
solar-type stars have migrated inwards from very large orbits. The need 
for this large radial interval of migration might then explain the 
presence of the brown dwarf desert (if massive brown dwarfs are generally 
not able to migrate as much because of the limited angular momentum of the 
circumstellar disk), and the lower eccentricities of the first population 
of planets (some of which would have migrated over small radial intervals 
and remained close to their initial circular orbits). Finally, the 
similarity in the eccentricity distributions of spectroscopic binaries 
does not seem to have any clear relation to this hypothesis.

\begin{acknowledgements} 
Eduard Masana is gratefully acknowledged for providing a code to perform 
Kolmogorov-Smirnov tests. We are grateful to Michael Norman and Scott 
Tremaine for insightful discussions. I.R. acknowledges support from the 
Spanish MEC through a Ram\'on y Cajal fellowship and project 
AYA2006-15623-C02. J.M. is supported by the Spanish MEC project 
AYA2005-09413-C02-01 with EC-FEDER funding and research project 
2005SGR00728 from Generalitat de Catalunya. J.M. thanks the Institute for 
Advanced Study in Princeton for their hospitality when this work was being 
done.
\end{acknowledgements}

\end{document}